\documentclass[amssymb,amsmath,superscriptaddress,footinbib]{revtex4}
\usepackage{amsmath,natbib,graphicx,subfigure,subeqnarray,fancyhdr,amsmath,multirow,color, mathrsfs,relsize}
\begin{document}

\newcommand{\rev}[1]{{\color{black}{#1}}}
\newcommand{\frev}[1]{{\color{black}{#1}}}

\title{Synchronized switch harvesting applied to piezoelectric flags.}

\author{Miguel Pi\~neirua}
\email{miguel.pineirua@gmail.com}
\affiliation{Unit\'e de M\'ecanique, ENSTA Paristech, Palaiseau, France}
\author{S\'ebastien Michelin}
\email{sebastien.michelin@ladhyx.polytechnique.fr}
\affiliation{LadHyX-D\'epartement de M\'ecanique, CNRS-Ecole Polytechnique, 91128 Palaiseau, France}
\author{Dejan Vasic}
\email{dejan.vasic@satie.ens-cachan.fr}
\affiliation{SATIE, ENS Cachan, Cachan, France}
\author{Olivier Doar\'e}
\email{olivier.doare@ensta-paristech.fr}
\date{\today}

\begin{abstract}
In this article the energy transfer between a flow and a fluttering piezoelectric plate is investigated. In particular, the benefits of the use of a Synchronized Switch Harvesting on Inductor (SSHI) circuit are studied. Both wind tunnel experiments and numerical simulations are conducted in order to analyse the influence of the switching process on the dynamics and the efficiency of the system. Numerical simulations consist of a weakly non-linear model of a plate in axial flow equipped with a single pair of piezoelectric patches, discretized using a Galerkin method where basis functions are the modes of the plate in vacuum. The discretized model is then integrated in time. The results presented in this paper show that a significant improvement of the harvested energy can be obtained using SSHI circuits compared to basic resistive circuits. It is also shown that for strongly coupled systems, the switching process inherent to he SSHI circuit has a significant impact on the dynamics of the flag,  which tends to  decrease the relative efficiency gain.
\end{abstract}
\maketitle

\section{Introduction}
In recent years, piezoelectric materials have received considerable attention in order to convert mechanical energy associated with ambient vibrations into electrical form, and power mobile and wireless microsystems~\cite{erturk2011}. Piezoelectric materials are also interesting to design new kinds of electric transformers \cite{cherif2014}. An important ingredient is however the existence of a vibration source. Hence, flow-induced vibrations have recently been considered as potential routes to generate spontaneous and self-sustained vibrations of  solid structures that can then be used to power an electric generator (piezoelectric or other): these include Vortex-Induced Vibrations \cite{bernitsas2008}, coupled mode flutter \cite{mckinney1981}, galloping \cite{barrero2010}, or flutter instability of flags \cite{tang2009b}. 

The latter is the focus of the present article: a plate covered by piezoelectric patches and placed in an \rev{axial} air- or water-flow of sufficient velocity undergoes large-amplitude self-sustained oscillations above a critical velocity threshold. \rev{These oscillations originate from an instability of the equilibrium position, resulting from the interaction of the flow forces with the solid’s inertia and rigidity. This instability is referred to as flag coupled mode flutter and has been the subject of numerous research papers in the last decades \cite{paidoussis2004,shelley2011}.} The idea of using fluttering piezoelectric plates to harvest flow \rev{kinetic} energy was born in the early 2000's \cite{allen2001} and numerous studies followed. Recently the influence of piezoelectric coupling on the stability properties and dynamics of the plate has been characterized \cite{doare2011b,dunnmon2011}, as well as the importance of mechanical-electrical synchronization \cite{michelin2013,xia2015} and of piezelectric patches' positions and dimensions \cite{Pineirua2015} in order to maximize the efficiency of the system. 

Electro-mechanical couplings of common piezoelectric materials are relatively low and an important challenge lies in the optimization of the output circuit to maximize the energy transfer. \rev{A broadly used technique in the optimization of mechanical to electrical power conversion from vibrating structures is the Synchronized Switch Harvesting on Inductor (SSHI) \cite{Guyomar2005}, which aims to synchronize the voltage and current in the output circuit so as to maximize the energy transfer.} There are two circuit topologies of SSHI: series and parallel.  In the latter, the circuit includes an active switch that briefly connects a parallel  inductive branch  when the displacement of the mechanical system is maximal, in order to reverse the voltage applied on the output load.   If the  system is properly tuned, this synchronization technique can  significantly increase the electrical power that can be extracted from \rev{the vibrations of the mechanical system \cite{Guyomar2005}}. The SSHI technique is widely studied in the domain of power electronics and various improvements have recently been proposed, such as self-powered SSHI \cite{lallart2008,chen2012}.  

\rev{The SSHI technique is generally applied in the optimization of the energy harnessing from vibrating piezoelectric structures excited by external driving forces.  Nevertheless, the application of this technique to  flow-induced vibrations of structures and its impact on the energy conversion efficiency of these  systems has not been studied yet.   Indeed, in the case of fluttering piezoelectric plates, the coupling between the  fluid-structure system  and the external circuit is of great importance in the overall efficiency of the system, as  the electric loading can dramatically influence the plate dynamics.}

\rev{Based on experiments and numerical simulations,} the objective of the present work is to characterize the influence of the SSHI parallel technique on the coupled fluid-solid-electric dynamics of a piezoelectric \rev{fluttering} flag, and to quantify the improvement on the energy harvesting efficiency it can provide. The paper is organized as follows: In section \ref{sec:modelingexp}, the piezoelectric flag system and SSHI circuits are presented and modeled, together with the experimental setup and numerical simulations techniques. In section \ref{sec:results}, the experimental and numerical data are presented for both weak and strong couplings of the mechanical and electrical systems, and the influence of the circuit's design on the harvesting efficiency is analyzed. Finally, section \ref{sec:conclusions} summarizes our results and offers some perspectives.

\section{Piezoelectric flag : modeling and experiments}
\label{sec:modelingexp}
\begin{figure*}[htb!]
\centering
 \begin{tabular}{l@{\hspace{0.05\textwidth}}l}
 (a)&(b)\\
 \includegraphics[width=0.45\textwidth]{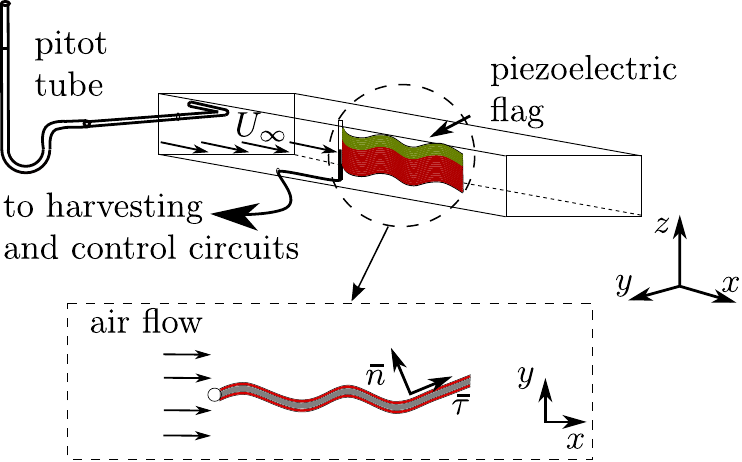}&
\includegraphics[width=0.45\textwidth]{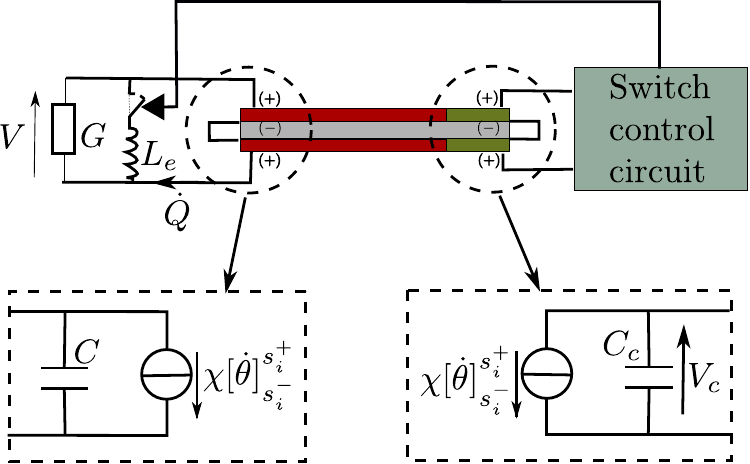}\\
&\\
(c)&(d)\\
 \includegraphics[width=0.4\textwidth]{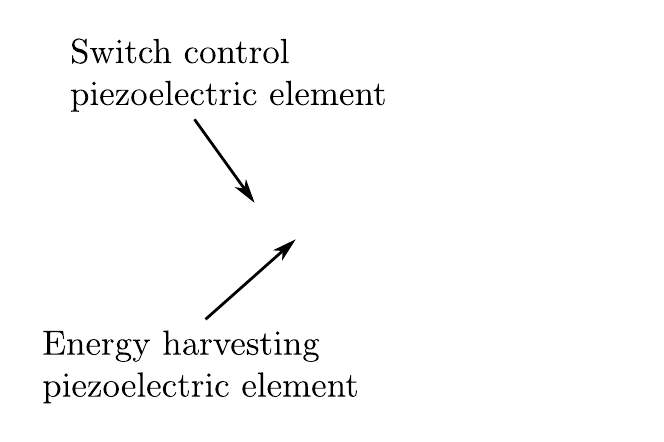}&
\includegraphics[width=0.45\textwidth]{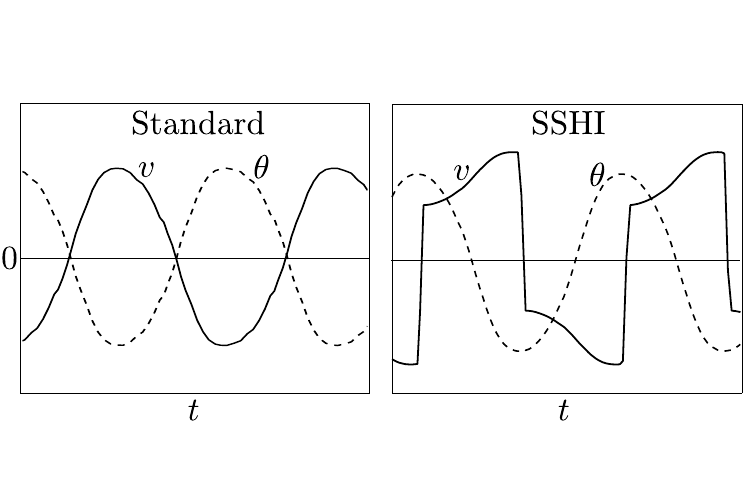}\\

 \end{tabular}
 \caption{\rev{(a) Description of the experimental setup} (b)  Connection diagram of the experimental setup.  The energy harvesting and switch control piezoelectric elements are shown in red and green respectively.  (c) Schematic of the multilayer composition of the piezoelectric flag.  (d) Temporal evolution of $v$ and $\theta$ for standard and SSHI configurations in open circuit conditions.}
\label{dispo_exp}
\end{figure*}

In this section we present a series of numerical simulations and experiments aiming to study the impact of the synchronized switching  on the energy harvesting efficiency of a piezoelectric flag. 
We focus here on a simple configuration consisting of a flexible piezoelectric plate  immersed in an axial wind flow (figure \ref{dispo_exp}-a).  The piezoelectric plate is connected to an electric circuit which consists of a resistive charge with a parallel inductive branch that can be switched on/off in order to inverse the piezoelectric voltage.  A physical model of the system used in the numerical simulations is presented  in the next subsection \ref{model}.  A detailed description of the experimental setup and materials are reported in subsection \ref{experiments}.  

\subsection{\label{model}Physical Model}

The piezoelectric flag considered here consists of an inextensible thin flexible plate of length $L$ and width $H$.  The plate is  clamped at its leading edge and placed in an axial wind flow of density $\rho$ and velocity $U_\infty$ (figure \ref{dispo_exp}-a). The entire plate is covered by two pairs of piezoelectric patches, and within each pair, the negative electrodes of the piezoelectric patches are shunted through the flag. The lineic mass and bending rigidity of the assembly are $\mu$ and $B$, respectively.

The output circuit is connected to the positive electrodes of the first piezoelectric pair, referred to in the following as harvesting assembly (see figure \ref{dispo_exp}-b). In the present approach, the energy dissipated in the resistance models the harvested energy. The positive electrodes of the second pair, referred to as the controlling patch assembly, are connected to the switch control circuit used to detect maxima of the plate's deflection by means of zero crossing velocity detection.

Beyond a critical wind velocity,  the plate starts to flap in a self sustained oscillatory regime. In the following, we consider only the two-dimensional motions of the plate in the $(x,y)$ plane.  The deformation of the piezoelectric patches induces a charge displacement between their electrodes and a net current within the resistive load. The total charge displacement $Q$ in the harvesting circuit is related to the plate's deformation and voltage $V$ by \cite{doare2011b},
\begin{equation}
Q=\chi\theta(L)+CV\;,
\label{eq_piezo}
\end{equation}
where $\theta(L)$ is the local orientation of the plate at the trailing edge with respect to the flow direction.  $C$ and $\chi$ are the capacitance and electromechanical coupling coefficients of the harvesting piezoelectric assembly, respectively~\cite{bisegna2006}. The voltage at the electrodes of the harvesting assembly induces a feedback mechanical forcing due to the inverse piezoelectric effects, in the form of an additional torque on the flag. Assuming that the flag itself follows an inextensible Euler--Bernoulli beam equation, its dynamics are described by
\begin{equation}
\rev{\mu\ddot{\mathbf{X}}(S,t)=\left[\mathcal{T}\boldsymbol\tau-M'\mathbf{n}\right]'+\mathbf{f}_\textrm{fluid},\qquad \mathbf{X}'(S,t)=\boldsymbol\tau},\label{eq:plaque}
\end{equation}
where the total torque within the flag is given by
\begin{equation}
M=B\theta'-\chi VF_v.
\end{equation}
\rev{In equation (\ref{eq:plaque}),} $(\boldsymbol\tau,\mathbf{n})$ are the unit tangent and normal vectors to the flag, $f_{fluid}$ is the loading exerted by the surrounding fluid and $\mathcal{T}$ the tension within the flag. $F_v$ is the polarization function of the piezoelectric element. In the present approach, $F_v(S)=\mathcal{H}(S)-\mathcal{H}(S-L)$, with $\mathcal{H}$ the Heaviside step function.  \rev{Derivatives with respect to the curvilinear coordinate ($S$) and time ($t$) are respectively noted $()'$ and~$\dot{()}$}. For slender flags, the fluid loading $\mathbf{f}_\textrm{fluid}$ is modeled locally in terms of the relative velocity of the flag to the background flow, using the Large Amplitude Elongated Body Theory~\cite{lighthill1971,candelier2011} combined with a resistive drag accounting for lateral flow separation~\cite{Eloy2012,singh2012a,michelin2013}.  The feedback of the controlling patch assembly is neglected: the second pair of piezoelectric patches acts only as a sensor that monitors  the deformation of the plate.

The harvesting circuit consists of a purely resistive load connected in parallel to an inductance $L_e$ associated with a switching device controlled by the second piezoelectric pair. The harvested energy is modeled here as the  energy dissipated in the resistive output. When the switching device is open (figure \ref{dispo_exp}-b), the current through the inductance $L_e$ is zero, and $Q$ and $V$ are simply related through Ohm's law: 
 \begin{equation}
 \dot{Q}+GV=0,
 \label{eq_circuit}
\end{equation}
where  $G$ is the conductivity of the load.  When the switching device is closed, Kirchhoff's law becomes  
\begin{equation}
 \ddot{Q}+G\dot{V}+\frac{V}{L_e}=0\label{eq:sshi}
\end{equation}
due to the presence of the inductance.

Using $L$, $L/U_\infty$, $U_\infty\sqrt{\mu L/C}$ and $U_\infty\sqrt{\mu C L}$ as characteristic length, time, voltage and charge respectively, the coupled dynamics of the flag and harvesting device can be completely described by five non-dimensional parameters, namely
\begin{align}
 U^*=LU_\infty\sqrt{\frac{\mu}{B}} ,& \qquad M^*=\frac{\rho  H L}{\mu} , \\
  \alpha=\chi\sqrt{\frac{L}{BC}} , \qquad \beta=&\frac{U_\infty C}{L G},\qquad \Gamma=\frac{L}{U_\infty \sqrt{L_e C}}.
\end{align}
The first four coefficients are the dimensionless flow velocity, fluid-solid mass ratio, electro-mechanical coupling factor, and tuning coefficient of the fluid-solid and electric systems respectively. The last coefficient, $\Gamma$, is a relative measure of the fundamental frequency of the L$_{\text{e}}$C circuit when the switch is closed with respect to the characteristic time scale of the flag dynamics. In the following, the L$_{\text{e}}$C reversal time scale will be much shorter than the flapping period and $\Gamma\gg 1$.

 In the present study the fluid-structure parameters will be fixed for the particular case of a flexible plate in air at constant speed $U_\infty$ just above the instability threshold ($U^*=14.5$, $M^*=0.67$).  A particular attention will be dedicated to the coupling factor $\alpha$ and tuning coefficient $\beta$, since they are indeed relevant in the evaluation of the active circuit system presented in this work.

 \subsection{Standard and SSHI energy harvesting}
In the following, two different energy harvesting methods are compared. The standard technique considers a purely resistive load without any inductance (i.e. the switch is not operated). In the SSHI technique, the inductive branch  is closed when $\theta(L)$ reaches its maximal and minimal values (see figure \ref{dispo_exp}-b).  When the inductive branch is closed, the inductance, capacitance and the resistance form a resonant circuit of short oscillating period $T\sim\sqrt{L_eC}$.  

Following \cite{Guyomar2005}, the switch is re-opened after one half oscillation period of the L$_{\text{e}}$C circuit resulting in an inversion of the voltage across the resistive load.  In order to achieve adequate shifting, the oscillating period of the L$_{\text{e}}$C circuit must be small compared to the oscillation period of the flapping flag, hence $\Gamma\gg 1$.  Typically, some losses occur during the voltage inversion process which are related to energy dissipation in the coil.  These losses are taken into account through an electric quality factor $Q_i$, such that the voltage after the short inverting process at $t=t_s$ is obtained as: 
\begin{equation}
V(t_s^+)=-V(t_s^-)e^{\frac{-\pi}{2Q_i}}.
\end{equation} 
The impact of $Q_i$ on the energy harvesting efficiency will be discussed in the following.

In both standard and SSHI harvesting techniques, the power dissipated in the resistive load is computed as the temporal average $P_e=\langle V^2 G \rangle$ over one oscillation period. The efficiency of the system, $\eta=P_e/P_f$, is defined as the ratio of the harvested electric power $P_e$ to the  fluid kinetic energy flux $P_f$ through the vertical cross-section occupied by the flag $P_f=1/2\rho AHU_\infty^3$, where $A$ is the oscillation amplitude of the trailing edge of the plate.

All other parameters being fixed, the harvesting efficiency vanishes in the open circuit ($G=0$) and short circuit ($G=\infty$) limits since $P_e\rightarrow 0$ in both cases. Therefore, $\eta$ is maximum for an intermediate value of $G$ (or $\beta$). In the following, $\beta$ is therefore varied to find the maximum dissipated power, which corresponds to a tuning of the flapping frequency to the characteristic timescale of the RC circuit consisting of the piezoelectric capacitance and output resistor \cite{michelin2013}. 

 \subsection{Numerical solution}
 The non-dimensional form of the dynamical equations for the plate and circuit are solved in the weakly nonlinear limit following the approach described in details in Ref.~\cite{Pineirua2015}. Equation~\eqref{eq:plaque} is first projected onto the $x$- and $y$-directions and the horizontal projection is used to eliminate the tension force from the $y$ projection using the inextensibility condition. Retaining only nonlinear terms up to $\mathcal{O}(y^3)$, a classical Galerkin decomposition is then employed: expanding the $y$-displacement on clamped-free beam eigenmodes, the beam equation is projected on the same set of eigenmodes.  After truncation to $N$ linear modes, the resulting coupled system of equations are integrated numerically using an explicit semi-adaptive fourth order Runge-Kutta method.
 
 \subsection{\label{experiments}Experimental setup}
The  piezoelectric plate's center core consists of a double-sided adhesive tape of length  $L=10$ cm, width $H=3$ cm and thickness $h_s=100$ $\mu$m. Both sides of the tape are  totally  covered by a piezoelectric PVDF film of length $L$  and width $0.75H$, for the energy harvesting piezoelectric patch and $0.25H$ for the controlling piezoelectric patch. In both cases the negative polarity of the piezoelectric film faces inwards. The thickness of the piezoelectric film is $50$ $\mu$m.  Figure \ref{dispo_exp}-c shows a schematic view of the plate's structure.  The piezoelectric PVDF films (Piezotech\textsuperscript{\textregistered}) are covered with an external Cr/Au layer which serves as the electrodes, connecting the piezoelectric patches to the energy harvesting and switching control circuits.  The piezoelectric flag geometric and physical characteristics are summarized in Tables \ref{table_piezo} and \ref{table_piezo2}.
\begin{table}[h]
\caption{Piezoelectric element characteristics}
\centering
\begin{tabular*}{\textwidth}{l @{\extracolsep{\fill}} c c}
&&\\
\hline
&harvesting element&zero detection element\\
\hline
length (cm)&10&10\\
width (cm)&2.25&0.75\\
thickness ($\mu$m)&50&50\\
\hline
\end{tabular*}
\label{table_piezo}
\end{table}
 \begin{table}[h]
\caption{Measured experimental parameters}
\centering
\begin{tabular*}{\textwidth}{l l l}
&&\\
\hline
Symbol&Description&Value (unit)\\
\hline
&&\\
\multicolumn{3}{l}{Fluid/Structure parameters}\\
&&\\
$\mu$&mass per unit length of the piezoelectric flag&7e-3 (kg/m)\\
$U_\infty$&windspeed&12.7 (m/s)\\
$L$&flag length&10 (cm)\\
$H$&flag width&3 (cm)\\
\multirow{2}{*}{$B$}&bending rigidity of the flag&\multirow{2}{*}{5.3e-5 (Nm$^2$)} \\
&(including piezoelectric elements)&\\
&&\\
\multicolumn{3}{l}{Electric parameters}\\
&&\\
$C$&circuit capacitance&58 (nF)\\
$L_e$&switching branch inductance&50 (mH)\\
$Q_i$&electric quality factor&1.9\\
$G$&load conductivity (variable)&1e-6 to 1e-4 (1/$\Omega$)\\
$\chi$&electro-mechanical coupling coefficient&3.6e-7 $\pm$ 3e-8 (C)\\
&&\\
\hline
\end{tabular*}
\label{table_piezo2}
\end{table}

 The piezoelectric plate is placed in a wind tunnel of $10\times 5$ cm section with transparent walls allowing for a visual access \cite{doare2011d}.  The overall piezoelectric coefficient $\alpha$ of the system is determined experimentally by measuring the generated output voltage  corresponding to the real-time plate deformation recovered via high speed camera acquisition.  For our system we obtain  a value of $\alpha=0.065\pm0.007$~\cite{xia2015b}. 
 
 The harvesting piezoelectric patch is connected to a purely resistive circuit of variable conductivity.  The voltage across the resistive load is used to determine the dissipated power $P_e$.  The controlling piezoelectric patch is connected to a high-impedance electronic circuit (i.e. the electric charge displacement in that second piezoelectric pair is negligible). When the voltage (or equivalently the trailing edge's deflection, Eq.~\eqref{eq_circuit}) reaches a maximum or a minimum value, the control circuit generates a pulse that closes the inductive branch in the harvesting circuit. The reopening of the switch is controlled manually by adjusting the triggering threshold in order to get only one half oscillating period  of the L$_{\text{e}}$C circuit.  
 
 The inductive branch consists of a $L_e=50$ mH inductor, which along with the piezoelectric capacitance $C=58$ nF, give $\Gamma\sim150$. In this way we assure that the oscillation period of the L$_{\text{e}}$C  circuit is much shorter than the flapping period of the flag.  The factor $Q_i$ related to dissipative losses in the inductor is measured directly from the experiments, by comparing the voltage values before and after the inversion.  In our experiments we obtain $Q_i\sim 1.9$, which is also chosen for consistency in the numerical simulations. 

\section{Results}
\label{sec:results}

Both standard and SSHI harvesting techniques are now compared using a combination of numerical simulations and experimental measurements.  In section~\ref{sec:weak}, the results of numerical simulations are compared to experiments in the case of a weakly-coupled system $\alpha \ll 1$.  Section~\ref{sec:strong} then presents numerical simulation results for a strongly-coupled system ($\alpha=0.5$) and analyses the impact of the coupling factor on the harvesting efficiency and the plate dynamics for a strongly-coupled system.

\subsection{Weakly coupled systems}\label{sec:weak}
\begin{figure}[h]
\centering
\includegraphics[width=0.7\textwidth]{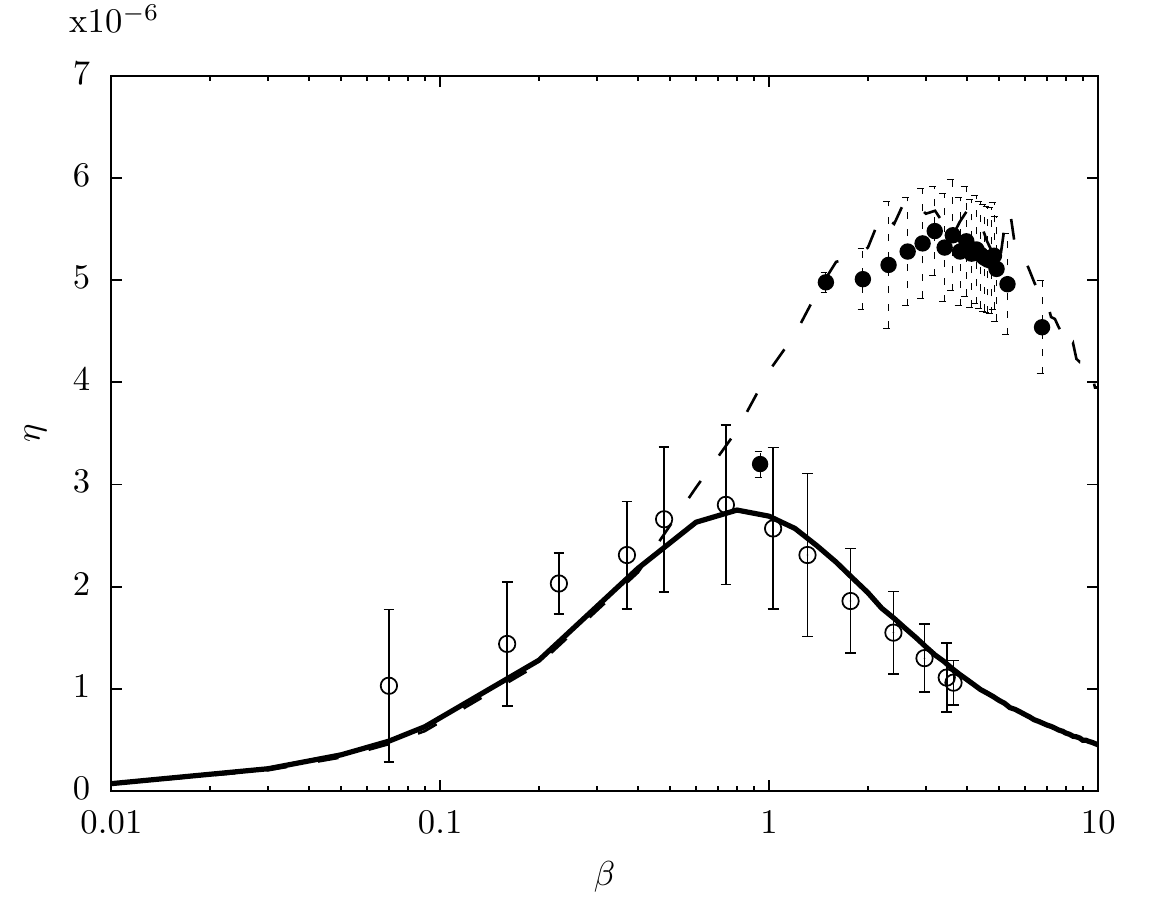}
\caption{Harvested energy efficiency as a function of the tuning parameter $\beta$ obtained in experimentally for the standard (open circles) and SSHI (filled circles) techniques, for $\alpha=0.065$, $Q_i=1.9$ and $\Gamma=180$. The corresponding results of numerical simulations are shown as solid and dashed lines.}
\label{beta_dep}
\end{figure}

\rev{In Figure \ref{beta_dep} we present the results of our experiments and numerical simulations comparing the standard and SSHI  techniques.}
As reported in previous works \cite{Guyomar2005} for other types of vibrating systems,  the SSHI harvesting technique substantially improves the harvesting efficiency of the system in the case of a weakly-coupled system (Figure \ref{beta_dep}).  An increase of the harvesting efficiency of about 100\% is observed in comparison with the standard harvesting technique. Also, the maximum efficiency is shifted to higher $\beta$ values with respect to the standard case.

  The experimental efficiency of the SSHI technique is  however limited by the low quality factor value of the inductive circuit $Q_i\sim1.9$.  In previous works on SSHI circuits \cite{Guyomar2005, chen2012}, the harvested power can be incremented up to 4 to 5 times with higher quality factors in the switching processes ($Q_i\sim5$).

Due to the limitations of the  electromechanical conversion factor of the PVDF film used in our experiments (which corresponds to a value of $\alpha=0.065$), our results remain in a weakly-coupled regime where the dynamics of the piezoelectric flag stay unperturbed by the electrical circuit charge. Nevertheless, based on the  good agreement between our experimental results and our numerical simulations, a numerical study of  the impact of high coupling factors on the harvesting efficiency of the system is presented in the following section.

\subsection{Strongly coupled systems}\label{sec:strong}

\begin{figure}[th!]
\begin{center}
\begin{tabular}{l l}
(a)&(b)\\
\includegraphics[width=0.45\textwidth]{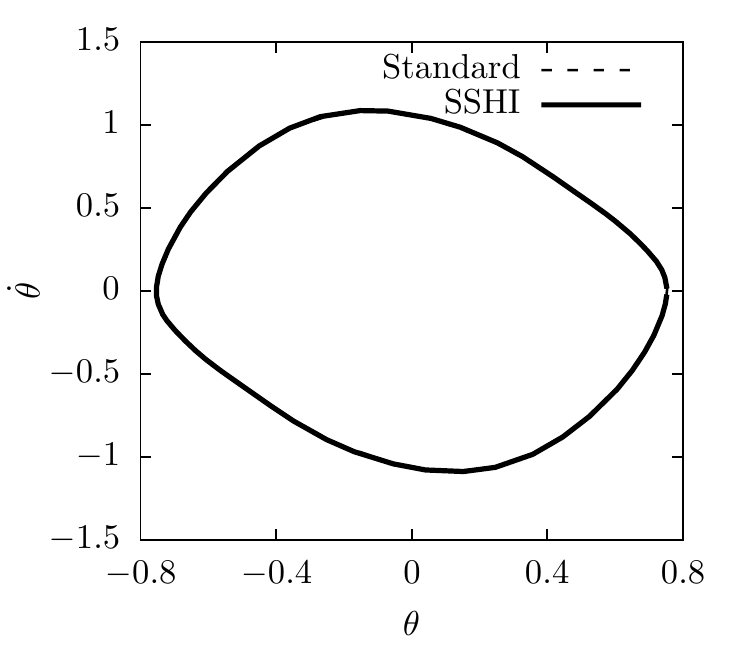}&
\includegraphics[width=0.45\textwidth]{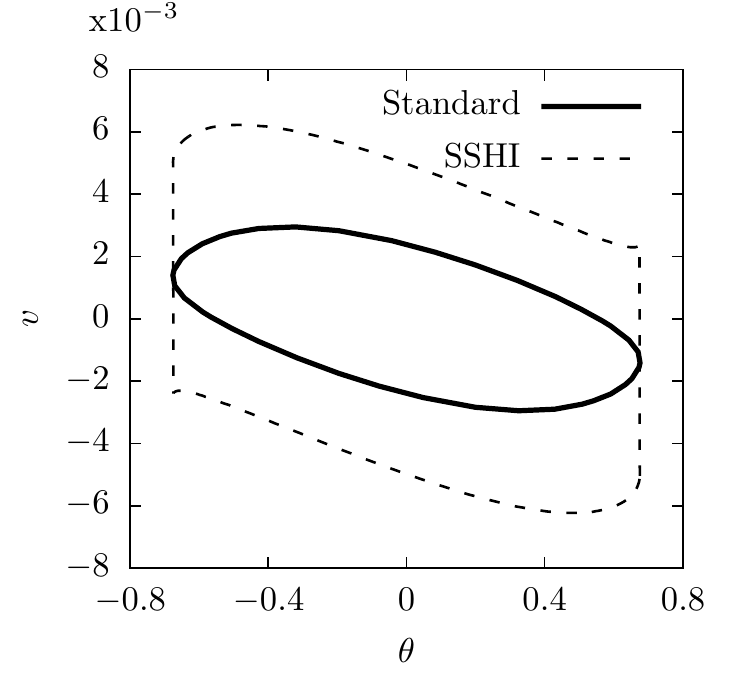}\\
(c)&(d)\\
\includegraphics[width=0.45\textwidth]{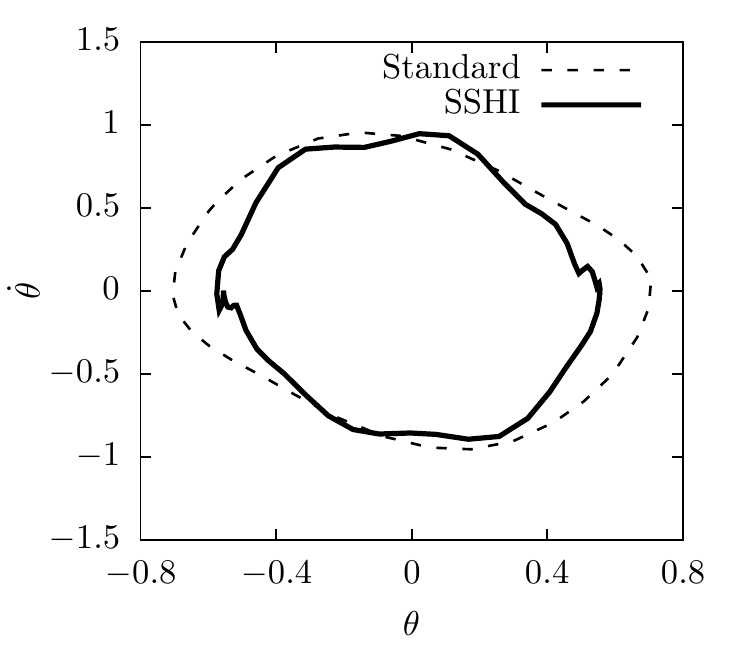}&
\includegraphics[width=0.45\textwidth]{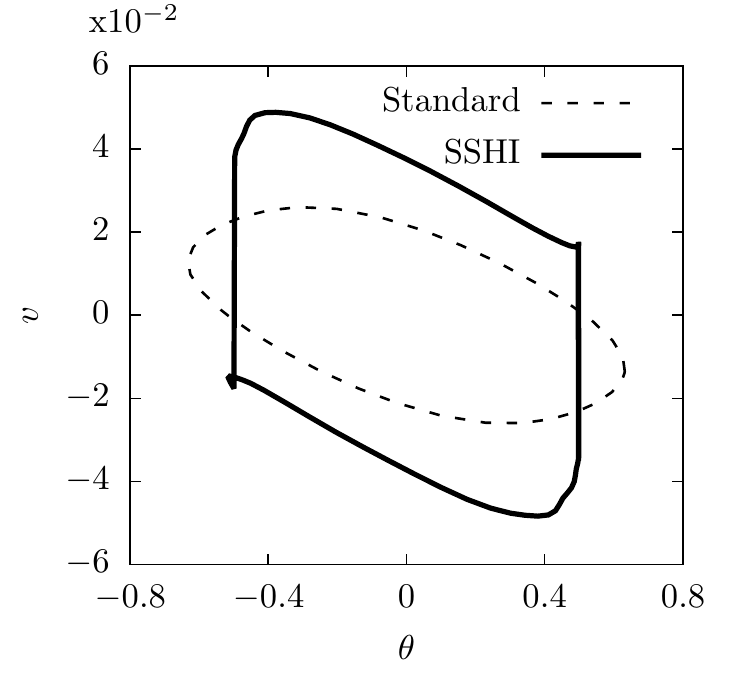}\\
\end{tabular}
\caption{Phase-space trajectory for the trailing edge orientation (left) and non-dimensional output voltage  cycle (right) for $\alpha=0.05$ (top) and $\alpha=0.5$ (bottom), and $Q_i=1.9$ and $\Gamma=180$.  $\beta$ is chosen to achieve a perfect tuning and maximum energy transfer: $\beta=0.9$ (std) and $\beta=4$ (SSHI). }
\label{alpha_dep}
\end{center}
\end{figure}
Large values of the electromechanical coupling factor generate larger charge displacements within the circuit for a given deformation, and thus enhance energy harvesting.  For strongly-coupled systems, the feedback piezoelectric torque on the flag also has an important impact on the fluid-solid-electric dynamics of the system, which has been showed to substantially affect the overall efficiency of the system~\cite{xia2015}.  Moreover, in the case of a strongly-coupled SSHI device, the sudden changes in the voltage polarity during the switching process can considerably perturb the plate dynamics.

\begin{figure}[t]
\begin{tabular}{l @{} l}
(a)&(b)\\
\includegraphics[width=0.45\textwidth]{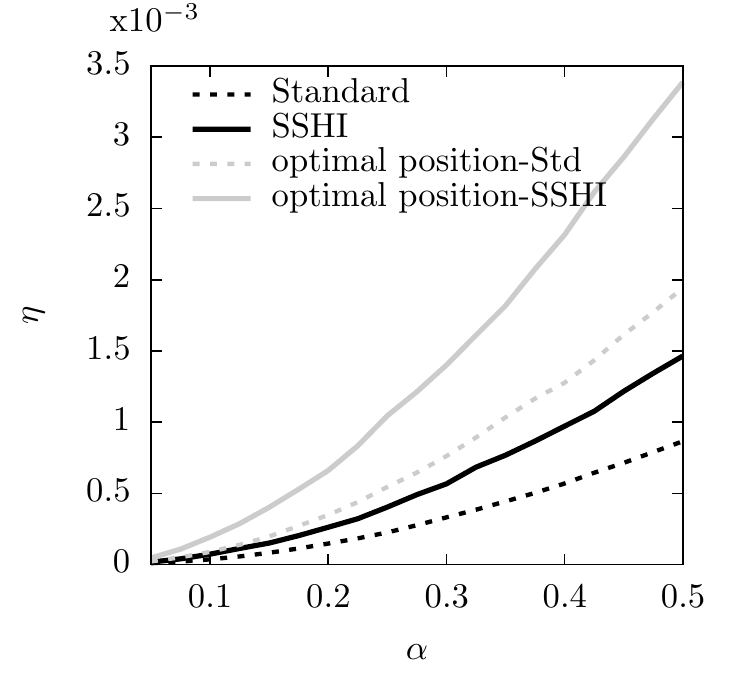}&
\includegraphics[width=0.45\textwidth]{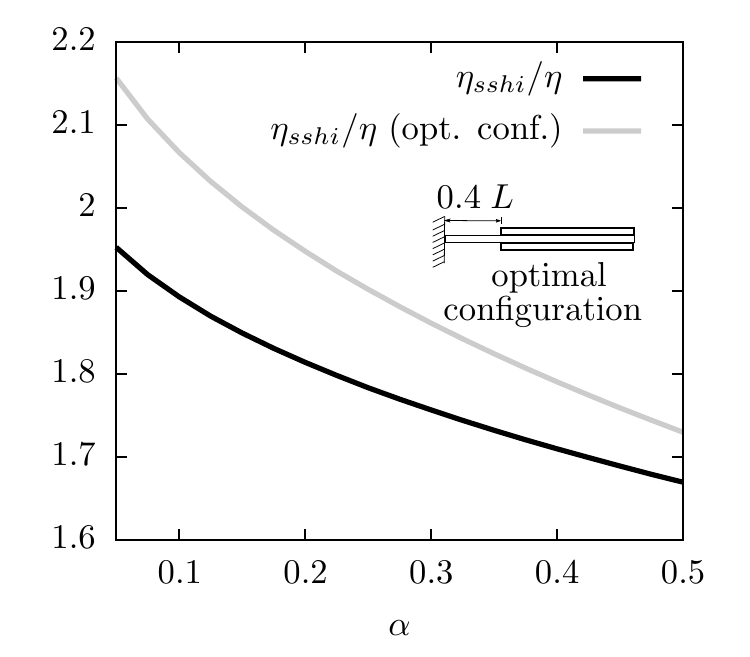}\\
\end{tabular}
\caption{(a) Energy harvesting efficiency as a function of the electro-mechanical coupling $\alpha$ for standard harvesting (solid) and SSHI harvesting (dashed), in the case of  a fully-covered piezoelectric plate (black) and a plate with an optimally-positioned piezoelectric element (gray)~\cite{Pineirua2015}. (b) Relative efficiency improvement introduced by the SSHI technique. $\Gamma$, $Q_i$ and $\beta$ are determined as in Figure~\ref{alpha_dep}. }
\label{effcompar_confopt}
\end{figure}

This impact is illustrated on Figure~\ref{alpha_dep}: for weak coupling, the dynamics of the plate remain unchanged between the standard and SSHI configurations (Figure~\ref{alpha_dep}a).  However, for strong coupling, the SSHI configuration significantly modifies the dynamics with a reduction of the flapping amplitude and the introduction of high-frequency components in the plate's vibrations near the switching event. These modifications of the dynamics also impact the efficiency. This is illustrated here on Figures  \ref{alpha_dep} (b) and (d) showing the recovered energy cycle for the same weakly and strongly coupled systems. Over one cycle, the energy transferred to the output circuit is equal to the enclosed area within the $\theta$-$v$ trajectory.  As expected, for strongly-coupled systems the amount of harvested energy per cycle is significantly larger than that of a weakly-coupled system.  However, the relative output power increase between the standard and SSHI configurations is smaller than for weakly-coupled systems.  In other words, for larger electro-mechanical coupling, the SSHI advantages are reduced (see figure \ref{effcompar_confopt}b). 

The core idea of SSHI is to synchronize the circuit's dynamics to the plate's motion to ensure a maximum transfer of energy to the electrical system. This ensures a maximum extraction of the mechanical energy of the flag, thereby reducing its flapping amplitude and the amount of mechanical energy available for conversion. The perturbation of the plate's dynamics due to the sudden switching of the inductive branch in the harvesting circuit (as shown in figure  \ref{alpha_dep}) could also explain the reduction of the SSHI advantages in strongly-coupled systems.

The positioning of the piezoelectric elements on the flag also plays an important role, as it has been shown in previous works~\cite{Pineirua2015}. In Figure~\ref{effcompar_confopt}-(a) we compare the efficiencies of a fully covered piezoelectric flag to those of a flag with an optimized piezoelectric element distribution.  We observe that the overall efficiency of the system is improved by an optimal positioning of the piezoelectric element for both standard and SSHI harvesting. Also, the optimal positioning of the piezoelectric elements improves the SSHI configuration effectiveness with respect to the standard configuration in about $10\%$ (Figure~\ref{effcompar_confopt}-(b)). 

\begin{figure}[h]
\centering
\includegraphics[width=0.73\textwidth]{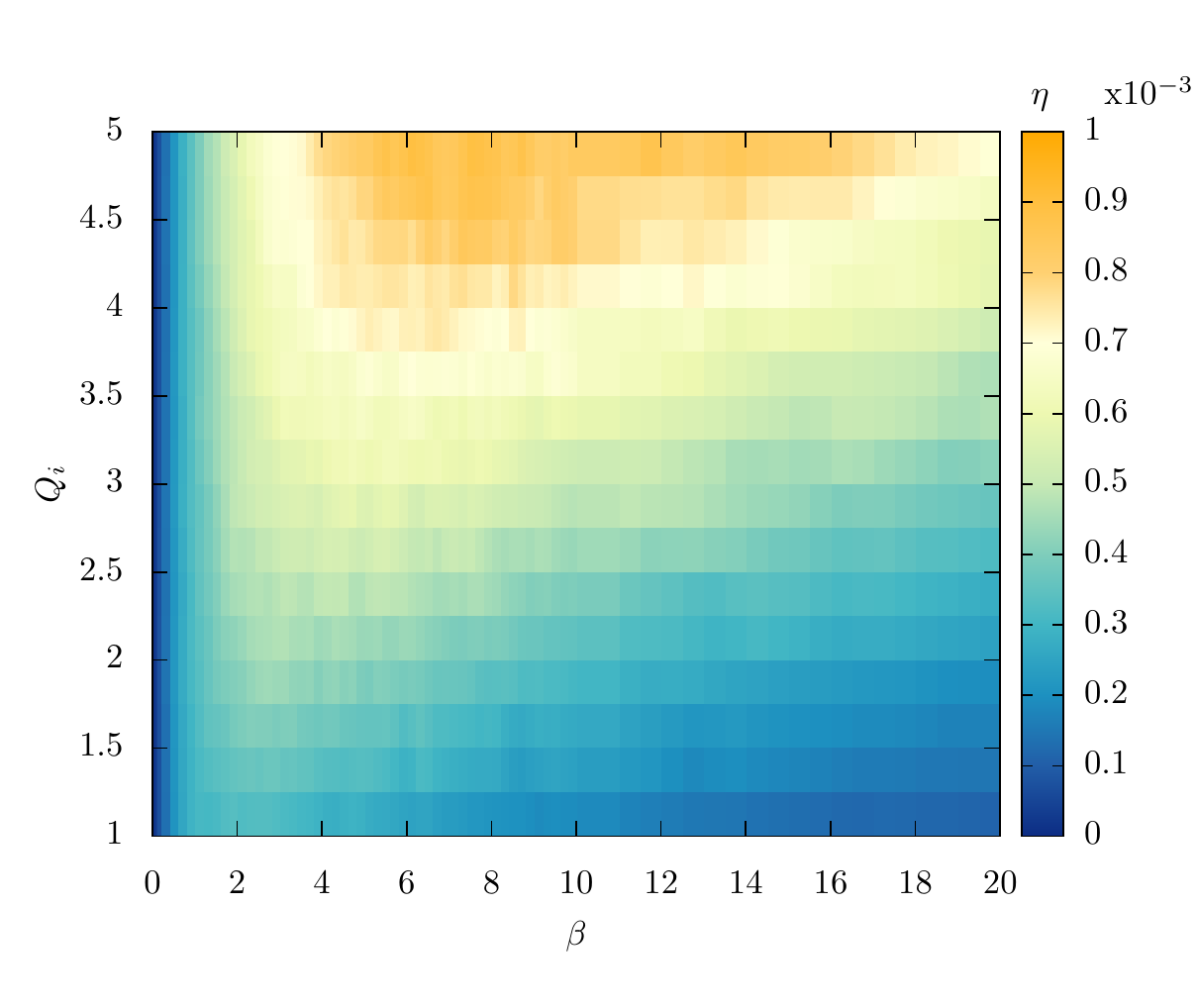}
\caption{Energy harvesting efficiency obtained for SSHI harvesting as a function of the quality factor $Q_i$, with $\alpha=0.25$ and $\Gamma=180$.}
\label{Qi_dep}
\end{figure}

Finally, we investigate the importance of the electric quality factor $Q_i$  in the efficiency of the SSHI technique. Confirming the results obtained in previous studies \cite{Guyomar2005,chen2012},  the harvested power  substantially increases with the quality factor (Figure~\ref{Qi_dep}).  We observe that the  tuning parameter $\beta$ maximizing the efficiency also increases with $Q_i$: the maximum efficiency is obtained for larger output resistances.  \rev{In standard energy harvesting configurations one generally obtains optimal $\beta$ values near unity \cite{michelin2013,Pineirua2015}.  Indeed, in such simple electrical configurations, power production is maximized when the characteristic timescales of the mechanical system ($L/U$) and of  the output circuit ($C/G$) are comparable.  However, in the SSHI case the synchronization between the mechanical and electrical systems is artificially forced, hence, the sense of $\beta$ as a measure of mechanical/electrical synchronization is lost. }

\section{Conclusion}\label{sec:conclusions}

The present work proposed an analysis of the energy harvesting efficiency of a piezoelectric flag using the SSHI technique, which has become quite popular to optimize energy harvesting from ambient vibrations. Using numerical simulations and experimental measurements, we showed that if the resistor is correctly tuned, SSHI allows for a doubling of the harvested power. A good agreement between experimental and numerical results was observed. The numerical model was further used to analyze the role of piezoelectric coupling in SSHI systems for strongly-coupled systems, i.e. when the flag's dynamics is impacted by the harvesting circuit. In particular, larger coupling coefficients and quality factors have been shown to enhance the harvesting efficiency. However, due to the piezoelectric feedback coupling on the flag, the relative increase of harvested power obtained for SSHI in comparison with the standard technique is reduced for strongly-coupled systems. 

Finally, using the guidelines proposed in our previous work \cite{Pineirua2015}, it was shown that the efficiency can again be increased by a factor 2 by optimizing the position of the harvesting electrodes along the flag's length. Hence these results show that significant improvement can be obtained if the geometries and electrical circuits of piezoelectric energy harvesters are carefully designed.

Future work should focus on the implementation of the SSHI technique on more complex systems, such as multiple piezoelectric element flags or multiple flag energy harvesting configurations.

\section*{Acknowledgements}
This work was supported by the ``Laboratoire d’Excellence'' LASIPS (project PIEZOFLAG) and by the French National Research Agency ANR (Grant ANR-2012-JS09-0017).  M.P. and  O.D. would like to thank Amalia P. Pellizzi and Octave D. Rouby for fruitful comments and discussions.


\end{document}